# Developing indicators on Open Access by combining evidence from diverse data sources

Thed van Leeuwen *, Ingeborg Meijer, Alfredo Yegros-Yegros and Rodrigo Costas

* Corresponding author: leeuwen@cwts.nl

CWTS,

Leiden University,

Wassenaarseweg 62a,

Leiden, the Netherlands

## Introduction

In the last couple of years, the role of Open Access (OA) publishing has become central in science management and research policy. In the UK and the Netherlands, national OA mandates require the scientific community to seriously consider publishing research outputs in OA forms. At the same time, other elements of Open Science are becoming also part of the debate, thus including not only publishing research outputs but also other related aspects of the chain of scientific knowledge production such as open peer review and open data. From a research management point of view, it is important to keep track of the progress made in the OA publishing debate. Until now, this has been quite problematic, given the fact that OA as a topic is hard to grasp by bibliometric methods, as most databases supporting bibliometric data lack exhaustive and accurate open access labeling of scientific publications (van Leeuwen et al, submitted). In this study, we present a methodology that systematically creates OA labels for large sets of publications processed in the Web of Science database. The methodology is based on the combination of diverse data sources that provide evidence of publications being OA.

## Our approach

Some reports have appeared over the years on OA publishing (Archambault et al, 2014). A common limitation in these studies lies in the methodology, which is usually based on random samplings in combination with a harvesting approaches to identify OA publications (i.e. publications are harvested online in the quest for free open versions in any form). This limitation has the important drawback that the results of these studies are difficult to reproduce and scale, as they are very much dependent on the sample selection, as well as on the moment and approach of the online harvesting (e.g. publications that appear as OA at the moment of harvesting may not appear as such at a later time, OA publications may disappear from sources that were seemingly offering them, etc.). The methodological approach that we propose mainly focuses at adding different OA labels to the complete in-house version of the Web of Science database (period 2009-2014), using various data sources to establish this OA status. Basic principles for this OA label are sustainability and legality. With sustainability we mean that it should, in principle, be possible to reproduce the OA labeling from the various sources used, again and again, in an open fashion, with a relatively limited risk of the source disappearing behind a pay-wall. The second aspect relates to the usage of data sources that represent legal OA evidence for publications, excluding rogue or illegal OA publications.

As main data sources we used:
- the DOAJ list (Directory of Open Access Journals) [https://doaj.org/],
- the ROAD list (Directory of Open Access scholarly Resources) [http://road.issn.org/],
- PMC (PubMed Central) [https://www.ncbi.nlm.nih.gov/pmc/],
- CrossRef [https://www.crossref.org/], and
- OpenAIRE [https://www.openaire.eu/]

All these sources fulfill the above mentioned requirements while other popular 'apparent' OA sources such as ResearchGate and SciHub fail to meet these two principle requirements. Thus, it is important to highlight here that our approach has a more policy and strategy perspective than a utilitarian one. In other words, our approach aims to inform the number and share of sustainable and legal OA publications (i.e. publications that have been published in OA journals or archived in official and legal repositories), instead of the mere identification of publications whose full text can be retrieved online (regardless the source or the legal status of the access to the publication).

**Sources of Open Access evidence**

The sources that were mentioned above were fully downloaded (as provided by the original sources) using their public Application Programming Interface (API). The obtained metadata has been parsed and incorporated into an SQL environment in the form of relational databases.

**DOAJ**

A first source we used is the DOAJ list of OA journals. This list was linked to the WoS database on the basis of the ISSN code available in both the DOAJ list as well as in the WoS database. This resulted in a recall of 1,685,420 publications labeled in the WoS as being OA.

**ROAD**

A next source used to add labels to the WoS database is the ROAD list. ROAD has been developed with the support of the UNESCO, and is related to ISSN International Centre . The list provides access to a subset of the ISSN Register. This subset comprises bibliographic records which describe scholarly resources in OA identified by an ISSN: journals, monographic series, conference proceedings and academic repositories. The linking of the ROAD list is based upon the ISSN code available in both the WoS as well as in the ROAD list. This resulted in total in 552,925 publications being labeled as OA.

**CrossRef**

A third source that was used to establish an Open Access is CrossRef was based upon the DOI's available in both systems. This lead to the establishment of in total 791,675 publications as being licensed as OA according to CrossRef.

**PubMed Central**

A forth source used is the PubMed Central database. This is done in two ways, the first based upon the DOI's available in both the PMC database as well as in the WoS database. This resulted in total in 1,950,147 publications being labeled as OA in the WoS environment (indicated as PMC-1). The second approach was based upon the PMID code (where PMID stands for PubMedID) in the PMC database as well as in the WoS database. This resulted in total in 2,285,145 publications being labeled as OA in the WoS database (indicated as PMC-2).

**OpenAIRE**

A fifth and final data source used to add OA labels to the WoS database is the openAIRE database. OpenAIRE is a European database that aggregates metadata on OA publications from multiple institutional repositories (mostly in Europe), including also thematic repositories such as arxiv.org. The matching is done in two different ways, based upon matching by using the DOI's or PMIDs available in both OpenAIRE and in WoS (indicated as OpenAIRE-1, resulting in 1,139,597 publications); and second, on a fuzzy matching principle of diverse bibliographic metadata both in WoS and OpenAIRE (including articles' titles, publication years and other bibliographic characteristics) (indicated as OpenAIRE-2, resulting in total in 1,266,042 publications) (the methodology is similar to the methodology for citation matching employed at CWTS – Olensky et al. 2016

**Validation of the matching procedure**

In order to provide some validation of the matching procedure, we applied three different validation steps into the process, of which we can discuss only the first one undertaken, given the length of this manuscript. Full description of this and the other two validation steps is given in the full paper of the study.

<u>Manual check the results of the matching procedure from linking WoS to OpenAIRE</u>. A random sample of 1,000 publications has been obtained from WoS. For these publications, we manually checked a number of aspects that we used in the automatic matching, such as years, first author names and titles of the publications. The matching procedure took into consideration 14% of the years on OpenAIRE to be different as compared to WoS, while the first authors on both ends were exactly the same for the 1,000 publications in the sample. Of the titles on both ends, nearly 50% contained differences, related to spaces and missing words in the OpenAIRE set (some 90% of the titles that differed. Finally, we compared the total outcome of the matching procedure:

   a. if the paper has been matched to only one OpenAIRE record (91%)
   b. if the paper has been matched to more than one OpenAIRE record (9%)
   c. whether the matched cases under a) are correct or not (we found 100% correct)
   d. if the matched cases under b) are in any of these possibilities: at least 1 correct (we found 100% correct).

**Results**

For the period 2009-2014, we selected publications from WoS, and added the matched publication sets as created above. This set from WoS contained in total 11,323,003 publications. Of that set, 7,738,419 are labeled as research articles of which in total 1,998,920 publications were labeled as OA in any of the sets created above (which is 26% overall).

The various approaches have led to the following trends in time, displayed in Figure 1 and 2. Figure 1 shows the share of OA labeled publications of any type in WoS, while Figure 2 shows the share of OA labeled papers in WoS classified as research articles.

The strong increase of the OpenAIRE-1 approach in 2014 as signaled in Figures 1 and 2 can be the result of changes in the databases used, probably in the number of DOIs included in openAIRE, as that is the key feature in the matching in that approach. Internationally, OA guidelines started to change in this year, for example the ERC launched new guidelines for OA publishing of results from ERC funded research grants (ERC, 2014). This suggest that the internal policies of the sources used to identify OA evidence need to be observed in the discussion of OA availability.

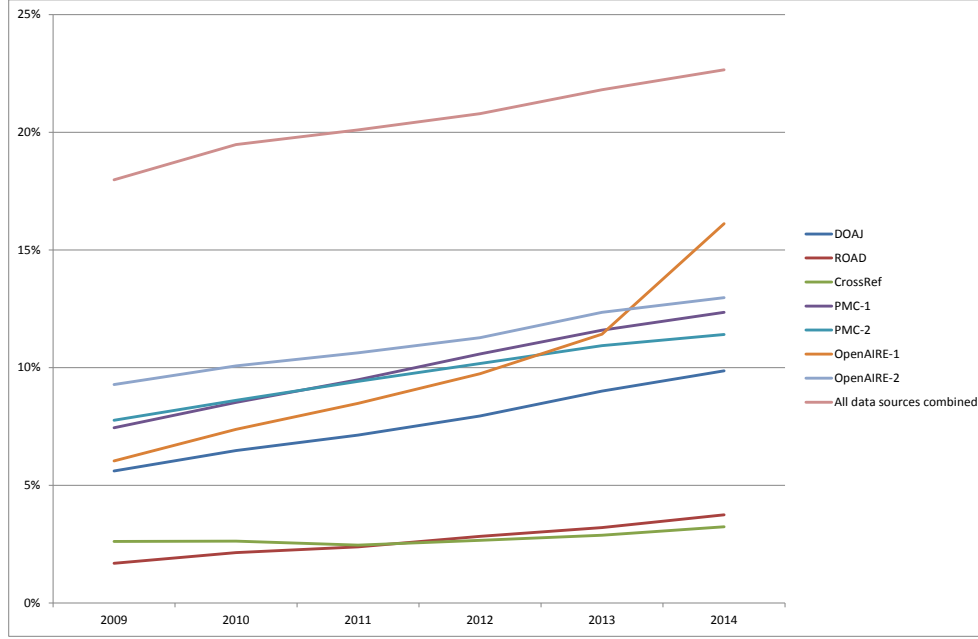

**Figure 1: Shares of Open Access labelled publications in WoS, 2009-2014**

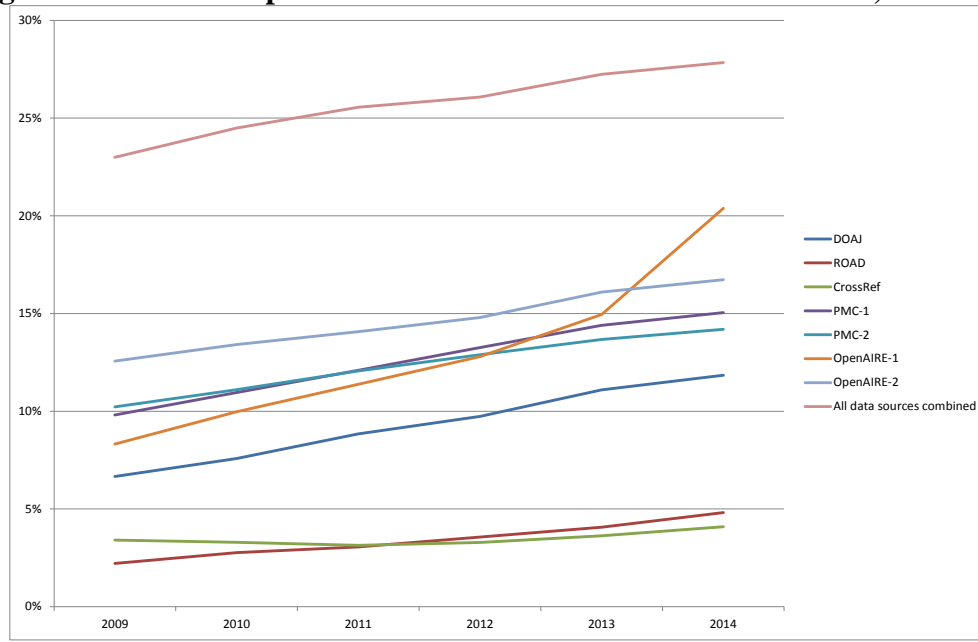

**Figure 2: Shares of Open Access labelled research articles in WoS, 2009-2014**

Overall, shares of Open Access availability are somewhat higher when we look at research articles only. Moreover, Figure 2 also contains the combined OA labeled shares of research articles, which mounts up to close to 30% of all the research articles in WoS. The figure also shows that not one single approach covers enough OA coverage in WoS to prioritize that method over others, the combination of the various methods is the strongest.

**The European landscape**

As a next step, we linked the publications labeled with OA status on the CWTS in-house WoS database, after which we could determine OA shares for the EU member states. The results are presented in Table 1. Observing the shares, we notice that the shares range between 20% (Latvia and Romania), to 37% (the Netherlands). Remarkable is also the relative high share of Great Britain (34%), a country which, like the Netherlands, have a strong drive towards open science, and national OA mandates into play.

**Table 1: Research articles of the EU countries, 2009-2014, and their OA shares**

| Country | All | OA | %OA | Gold | %OA Gold | Green | %OA Green |
|---|---|---|---|---|---|---|---|
| AUSTRIA | 73576 | 23246 | 32% | 6423 | 28% | 16823 | 72% |
| BELGIUM | 107808 | 36714 | 34% | 9368 | 26% | 27346 | 74% |
| BULGARIA | 13358 | 3078 | 23% | 949 | 31% | 2129 | 69% |
| CYPRUS | 5088 | 1365 | 27% | 445 | 33% | 920 | 67% |
| CZECH REPUBLIC | 59162 | 16818 | 28% | 7734 | 46% | 9084 | 54% |
| DENMARK | 79619 | 25674 | 32% | 8573 | 33% | 17101 | 67% |
| ESTONIA | 9135 | 2809 | 31% | 1199 | 43% | 1610 | 57% |
| FINLAND | 63830 | 19086 | 30% | 6681 | 35% | 12405 | 65% |
| FRANCE | 387911 | 128271 | 33% | 30524 | 24% | 97747 | 76% |
| GERMANY | 548407 | 166323 | 30% | 47065 | 28% | 119258 | 72% |
| GREAT BRITAIN | 604747 | 206234 | 34% | 51646 | 25% | 154588 | 75% |
| GREECE | 59680 | 13656 | 23% | 5015 | 37% | 8641 | 63% |
| HUNGARY | 35236 | 10910 | 31% | 3233 | 30% | 7677 | 70% |
| IRELAND | 40599 | 13912 | 34% | 3181 | 23% | 10731 | 77% |
| ITALY | 321083 | 87769 | 27% | 28121 | 32% | 59648 | 68% |
| LATVIA | 3234 | 640 | 20% | 262 | 41% | 378 | 59% |
| LITHUANIA | 12144 | 2811 | 23% | 1657 | 59% | 1154 | 41% |
| LUXEMBOURG | 4140 | 1335 | 32% | 469 | 35% | 866 | 65% |
| MALTA | 1045 | 242 | 23% | 111 | 46% | 131 | 54% |
| NETHERLANDS | 193334 | 71019 | 37% | 18545 | 26% | 52474 | 74% |
| POLAND | 126549 | 34258 | 27% | 15617 | 46% | 18641 | 54% |
| PORTUGAL | 64639 | 20508 | 32% | 6636 | 32% | 13872 | 68% |
| ROMANIA | 42943 | 8798 | 20% | 4568 | 52% | 4230 | 48% |
| SLOVAKIA | 18381 | 5174 | 28% | 2500 | 48% | 2674 | 52% |
| SLOVENIA | 21503 | 6967 | 32% | 3411 | 49% | 3556 | 51% |
| SPAIN | 293156 | 93994 | 32% | 35362 | 38% | 58632 | 62% |
| SWEDEN | 128574 | 43491 | 34% | 14381 | 33% | 29110 | 67% |

In Figure 3, we present the OA covered share of the national outputs across the European landscape. In red we indicate all the countries with over 25% of their national output being published as OA, and the various colors of red show variations in the above 25% realm, while the blue colored countries have less than 25% of their national output published as OA output. Again, darker colors blue indicate higher shares of OA output in the national outputs as compared to lighter shades of blue.

**Figure 3: Shares of Open Access labeled publications national outputs of European countries (in WoS, 2009-2014 (red >25%, blue =< 25%))**

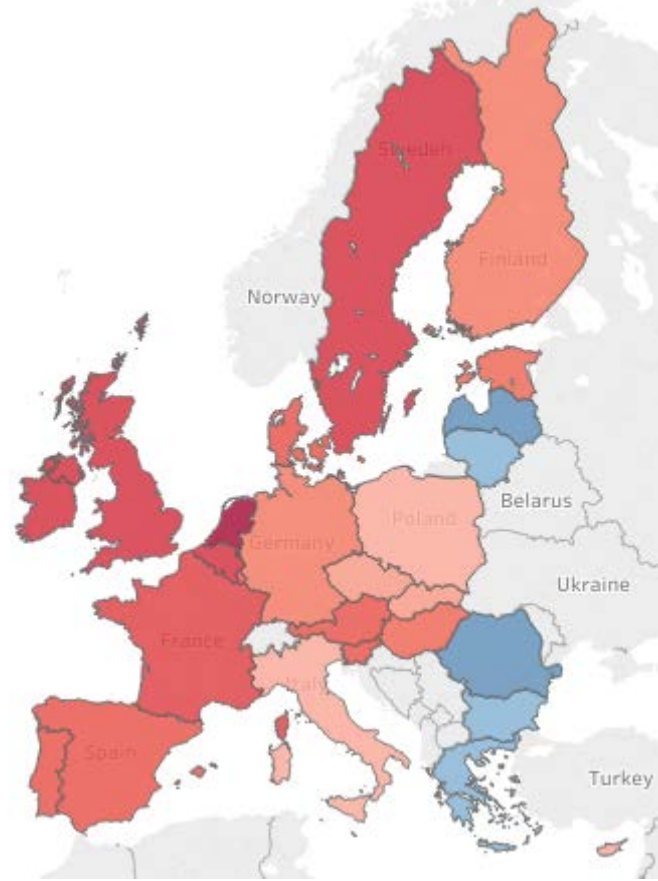

In Table 2, we present the numbers of publications covered by any of the sources we used for this study, across the various European countries. We present percentages OA (per source) of the total production , as that makes very clear that the conclusion drawn on the basis of the overall situation, that not one single source for attributing OA evidence is enough, is supported by the findings on the country level. Table 2 clearly shows that across the countries selected, the degree of OA attribution varies strongly, while for each country separately, none of the OA sources covers all of the OA attributions alone. It is important to notice the overlap among the different sources, as the sum of the individual shares outperforms the total OA share of the country. For each source we have highlighted in green the three countries with the largest shares and in red the three countries with the lowest ones for each of the sources.

**Table 2: Research articles of the EU countries, 2009-2014, and their OA data sources (in %)**

| Country | OA publications | %OA overall | % DOAJ | % ROAD | % CrossRef | % PMC-1 | % PMC-2 | % OpenAIRE-1 | % OpenAIRE-2 |
|---|---|---|---|---|---|---|---|---|---|
| AUSTRIA | 23246 | 32% | 8% | 4% | 7% | 16% | 14% | 22% | 20% |
| BELGIUM | 36714 | 34% | 8% | 4% | 4% | 12% | 12% | 23% | 23% |

| Country | OA publications | %OA overall | % DOAJ | % ROAD | % CrossRef | % PMC-1 | % PMC-2 | % OpenAIRE-1 | % OpenAIRE-2 |
|---|---|---|---|---|---|---|---|---|---|
| BULGARIA | 3078 | 23% | 7% | 2% | 6% | 4% | 3% | 15% | 13% |
| CYPRUS | 1365 | 27% | 8% | 3% | 6% | 6% | 6% | 20% | 15% |
| CZECH REPUBLIC | 16818 | 28% | 12% | 3% | 5% | 7% | 7% | 16% | 14% |
| DENMARK | 25674 | 32% | 10% | 5% | 4% | 16% | 16% | 23% | 21% |
| ESTONIA | 2809 | 31% | 12% | 3% | 4% | 11% | 11% | 22% | 17% |
| FINLAND | 19086 | 30% | 10% | 4% | 5% | 13% | 12% | 20% | 18% |
| FRANCE | 128271 | 33% | 7% | 3% | 6% | 11% | 10% | 23% | 22% |
| GERMANY | 166323 | 30% | 8% | 4% | 6% | 12% | 12% | 21% | 20% |
| GREAT BRITAIN | 206234 | 34% | 8% | 4% | 5% | 17% | 16% | 25% | 23% |
| GREECE | 13656 | 23% | 7% | 3% | 4% | 8% | 8% | 15% | 13% |
| HUNGARY | 10910 | 31% | 9% | 3% | 6% | 10% | 9% | 21% | 18% |
| IRELAND | 13912 | 34% | 7% | 4% | 4% | 12% | 12% | 25% | 21% |
| ITALY | 87769 | 27% | 8% | 3% | 5% | 10% | 10% | 19% | 16% |
| LATVIA | 640 | 20% | 8% | 3% | 4% | 7% | 6% | 12% | 10% |
| LITHUANIA | 2811 | 23% | 13% | 3% | 3% | 5% | 4% | 12% | 8% |
| LUXEMBOURG | 1335 | 32% | 11% | 5% | 3% | 12% | 12% | 22% | 21% |
| MALTA | 242 | 23% | 10% | 3% | 4% | 7% | 6% | 13% | 12% |
| NETHERLANDS | 71019 | 37% | 9% | 4% | 4% | 17% | 16% | 24% | 24% |
| POLAND | 34258 | 27% | 11% | 3% | 5% | 7% | 7% | 15% | 13% |
| PORTUGAL | 20508 | 32% | 10% | 4% | 4% | 8% | 8% | 20% | 20% |
| ROMANIA | 8798 | 20% | 10% | 2% | 4% | 3% | 3% | 10% | 8% |
| SLOVAKIA | 5174 | 28% | 13% | 2% | 6% | 5% | 5% | 14% | 12% |
| SLOVENIA | 6967 | 32% | 15% | 3% | 6% | 7% | 7% | 16% | 15% |
| SPAIN | 93994 | 32% | 11% | 4% | 4% | 9% | 9% | 22% | 14% |
| SWEDEN | 43491 | 34% | 10% | 5% | 5% | 17% | 16% | 24% | 22% |

## Conclusions and discussion

Given the fact that OA labels are not easily available across main sources for bibliometric studies, such as WoS and Scopus, and science policy is very alert on monitoring the developments of the change towards more open access availability of scientific knowledge, a clear urgency exists to construct OA labels ourselves. Therefore, we carefully selected a number of sources, and excluding others. Exclusion of sources was driven by two characteristics of the data sources: they had to be legal and sustainable. As such Research Gate, Academia.edu or SciHub were excluded, as these sources do not fully comply to these criteria. Recent developments such as 'oadoi' (https://oadoi.org/), which is the source for http://unpaywall.org, follow an approach very similar to the one presented here in order to identify OA versions of publications. In the future these sources will be also studied in order to be incorporated in the calculations.

Next, we described the developments of OA publishing in the context of the EU research environment. The European Union has initiated various steps to improve OA publishing (e.g. OA pilots for FP-7 funded research projects, OA Mandates, etc.), while some of the EU countries have actively stimulated OA publishing, by implementing OA mandates.

In this short paper, we have only concentrated on presenting our approach and studying OA publishing (vs. non OA) in Europe. In the full version of the paper we will include the urgent question on the effect of impacts related to OA publishing. What has become apparent in this study so far, is that information on OA publishing is rather dispersed, which means that our results clearly show that not one single source is sufficiently 'strong' to be taken into account for OA labeling of publications. It is in the combination of various sources what creates the most complete picture. Further research will focus on making the distinctions between Green and Gold OA, as well as hybrid and other forms of OA. All in all, our methodology seems to provide a systematic and quite complete perspective on the presence of OA publications in the European context.

# References

Archambault, É., Amyot, D., Deschamps, P., Nicol, A., Provencher, F., Rebout, L., and Roberge, G. Proportion of Open Access Papers Published in Peer-Reviewed Journals at the European and World Levels—1996–2013, Report to the European Commission, April 2014

ERC, 2014, "Open Access Guidelines for research results funded by the ERC", (https://erc.europa.eu/sites/default/files/document/file/ERC_Open_Access_Guidelines-revised_2014.pdf)

Van Leeuwen, T.N., C.T. Tatum, and P.F. Wouters, Exploring possibilities to use bibliometric data to study Open Access publishing on the national level. Submitted to *Journal of the Association for Information Science and Technology*

Olensky, M., Schmidt, M., & Van Eck, N.J. (2016). Evaluation of the Citation Matching Algorithms of CWTS and iFQ in Comparison to the Web of Science. *Journal of the Association for Information Science and Technology*, 67(10), 2550-2564. doi:10.1002/asi.23590.